\documentclass[
 reprint,
superscriptaddress,
amsmath,
amssymb,
aps,
reprint,
superscriptaddress
]{revtex4-2}

\usepackage[T1]{fontenc}
\usepackage{graphicx}
\usepackage{dcolumn}
\usepackage{bm}
\usepackage{braket}
\usepackage{mathptmx}
\usepackage[normalem]{ulem}
\usepackage{color}

\begin{document}
\title{Efficient backcasting search for optical quantum state synthesis}

\author{Kosuke Fukui}
\affiliation{%
Department of Applied Physics, School of Engineering, The University of Tokyo,\\
7-3-1 Hongo, Bunkyo-ku, Tokyo 113-8656, Japan}
\author{Shuntaro Takeda}
\affiliation{%
Department of Applied Physics, School of Engineering, The University of Tokyo,\\
7-3-1 Hongo, Bunkyo-ku, Tokyo 113-8656, Japan}
\author{Mamoru Endo}
\affiliation{%
Department of Applied Physics, School of Engineering, The University of Tokyo,\\
7-3-1 Hongo, Bunkyo-ku, Tokyo 113-8656, Japan}
\author{Warit Asavanant}
\affiliation{%
Department of Applied Physics, School of Engineering, The University of Tokyo,\\
7-3-1 Hongo, Bunkyo-ku, Tokyo 113-8656, Japan}
\author{Jun-ichi Yoshikawa}
\affiliation{%
Department of Applied Physics, School of Engineering, The University of Tokyo,\\
7-3-1 Hongo, Bunkyo-ku, Tokyo 113-8656, Japan}
\author{Peter van Loock} 
\affiliation{
Institute of Physics, Johannes Gutenberg-Universit\"{a}t Mainz, Staudingerweg 7, 55128 Mainz, Germany}
\author{Akira Furusawa} 
\affiliation{%
Department of Applied Physics, School of Engineering, The University of Tokyo,\\
7-3-1 Hongo, Bunkyo-ku, Tokyo 113-8656, Japan}
\affiliation{%
Optical Quantum Computing Research Team, RIKEN Center for Quantum Computing,\\
2-1 Hirosawa, Wako, Saitama 351-0198, Japan}

\begin{abstract}
Non-Gaussian states are essential for many optical quantum technologies.
The so-called optical quantum state synthesizer (OQSS), consisting of Gaussian input states, linear optics, and photon-number resolving detectors, is a promising method for non-Gaussian state preparation.
However, an inevitable and crucial problem is the complexity of the numerical simulation of the state preparation on a classical computer. 
This problem makes it very challenging to generate important non-Gaussian states required for advanced quantum information processing. Thus, an efficient method to design OQSS circuits is highly desirable. To circumvent the problem,
we offer a scheme employing a backcasting approach, where the circuit of OQSS is divided into some sublayers, and we simulate the OQSS backwards from final to first layers. 
Moreover, our results show that the detected photon number by each detector is at most 2, which can significantly reduce the requirements for the photon-number resolving detector.
By virtue of the potential for the preparation of a wide variety of non-Gaussian states, the proposed OQSS can be a key ingredient in general optical quantum information processing.
\end{abstract} 

\maketitle

{\it Introduction.}---
A non-Gaussian state is a key ingredient for quantum information processing, since the non-Gaussian feature is crucial for achieving universal and fault-tolerant quantum computation (FTQC) with optics~\cite{lloyd1999quantum,gottesman2001encoding,mari2012positive, veitch2012negative, menicucci2014fault,fukui2018high,baragiola2019all,pantaleoni2020modular,walshe2020continuous,pantaleoni2021subsystem}. In addition, it is essential for many applications~\cite{bachor2004guide} such as entanglement distillation~\cite{eisert2002distilling,dong2008experimental,hage2008preparation,takahashi2010entanglement}, bosonic error-correcting codes~\cite{cochrane1999macroscopically,leghtas2013hardware,mirrahimi2014dynamically,michael2016new,bergmann2016quantum,grimsmo2020quantum,albert2020robust}, quantum communication~\cite{namiki2014gaussian,sabapathy2017non,dias2017quantum,rozpkedek2021quantum,fukui2021all}, quantum metrology~\cite{huver2008entangled,anisimov2010quantum}, cloning~\cite{cerf2005non}.
The non-Gaussian state preparation is therefore a major effort in quantum information processing, and it has been extensively studied both theoretically~\cite{yurke1986generating,dakna1997generating,sasaki2006multimode,glancy2008methods,weigand2018breeding,eaton2019non,asavanant2021wave,takase2021generation}
and experimentally~\cite{wenger2004non,neergaard2006generation,ourjoumtsev2006generating,ourjoumtsev2007generation,gerrits2010generation,vlastakis2013deterministically,yukawa2013generating,yoshikawa2018heralded,fluhmann2019encoding,campagne2020quantum,grimm2020stabilization}.
Reviews on non-Gaussian state preparation can be found in references~\cite{lvovsky2020production,walschaers2021non}.

The non-Gaussian state preparation using Gaussian inputs, linear optics, and photon-number resolving (PNR) detectors is a promising way in optics~\cite{su2019conversion,
sabapathy2019production}, which we refer to as {\it optical quantum state synthesizer (OQSS).}
The striking feature of the OQSS is the ability to prepare, in principle, any superposition with an arbitrary pattern of Fock-state coefficients, which means that the OQSS can prepare an arbitrary single-mode non-Gaussian state.
In the OQSS, we need to simulate the state preparation and optimize the circuit parameters with a classical computer so that OQSS prepares the desired pattern of Fock-state coefficients.
However, an inevitable and crucial problem is the complexity of the numerical simulation of the state preparation on a classical computer: the computational time scales exponentially with the number of input modes for the circuit of OQSS.
This limits the number of patterns of coefficients we can optimize.
The complexity mainly comes from the calculation of a $loop$ $hafnian$ which is contained in the class of \#P-complete problems~\cite{valiant1979complexity,quesada2019franck}, where the complexity scales as an exponential time with the number of Gaussian inputs.
The complexity of the calculation of a hafnian is used for achieving a quantum supremacy~\cite{arute2019quantum} over a classical computer. In fact, a quantum supremacy has been demonstrated by a protocol using linear optics~\cite{zhong2020quantum}, commonly referred to as Gaussian boson sampling~\cite{aaronson2011computational}.

The complexity problem limits the set of non-Gaussian states available for advanced quantum information processing. Thus, a systematic and efficient method to find an experimental setup for a non-Gaussian target state preparation would be highly beneficial. In this letter, we develop an efficient technique to design OQSS circuits, effectively circumventing the complexity problem by applying ideas of a backcasting approach to the OQSS circuit that is decomposed into smaller, more tractable sublayers. As an important example and application, we numerically show that our OQSS method can simulate the preparation of the Gottesman-Kitaev-Preskill (GKP) qubit with a fidelity sufficient for FTQC~\cite{gottesman2001encoding} in polynomial time for the number of patterns of the Fock-state coefficients.

\begin{figure}[t]
 \centering \includegraphics[angle=0, scale=1]{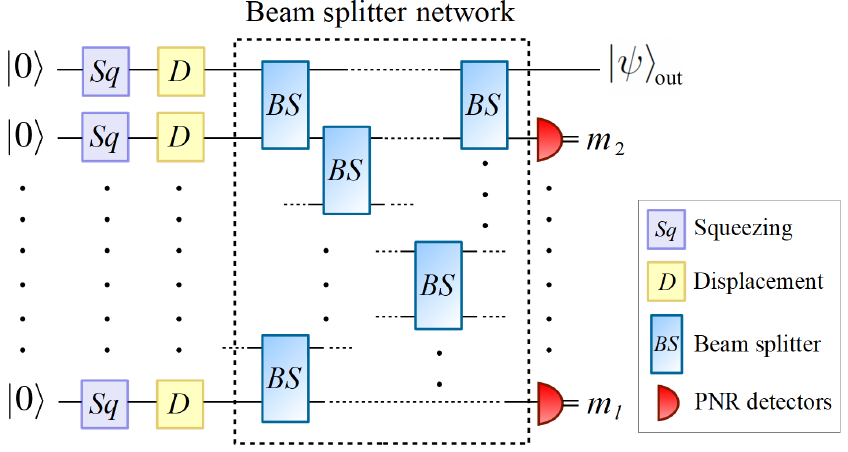} 
\caption{Schematic diagram for the OQSS consisting of linear optics and PNR detectors. 
The output state $\ket{\psi}_{\rm out}$ is generated so as to be close to the target non-Gaussian state.}
\label{fig1}
\end{figure}

{\it Optical quantum state synthesizer.}---
Figure \ref{fig1} shows the schematic diagram for the OQSS, where $l$ input vacuums are initially squeezed and displaced, then combined at a beam-splitter network, and finally all modes except that for the output state are measured by PNR detectors. 
Depending on the pattern of the detected photon numbers and circuit parameters for linear optics, the output $\ket{\psi}_{\rm out}$ is prepared as
\begin{equation}
\ket{\psi}_{\rm out}\approx U\sum_{i=0}^{n_{\rm max}}\frac{c_{i}}{N}\ket{i}, \label{eqfock}
\end{equation}
where $n_{\rm max}$, $c_{i}$, and $N$ correspond to the truncated photon number in the Fock basis, the coefficient of the Fock state $\ket{i}$ for the eigenvalue $i$, and a normalization factor, respectively, and $U$ is composed of Gaussian operations for a single mode (e.g., squeezing, displacement and rotation)~{\cite{su2019conversion, noteStellar}}.
The truncated photon-number $n_{\rm max}$ is given by
$n_{\rm max}=\sum_{i=2}^{l}{m_{i}},$
and coefficients $c_{i}$ depend on $m_{i}$ and the circuit parameters such as a transmittivity of a beam splitter, the amount of a squeezing, and the amount of a displacement~\cite{su2019conversion}.
These parameters are optimized by an optimization algorithm running on a classical computer so that $\ket{\psi}_{\rm out}$ becomes close to the target non-Gaussian state. 
In order to prepare an arbitrary non-Gausssian state using OQSS with $n_{max}$, we need to optimize $n_{\rm max}+1$ coefficients of the generated state for the Fock basis states from $\ket{0}$ to $\ket{n_{\rm max}}$. 
The number of independent coefficients, which we can access for optimization, has been conjectured~\cite{su2019conversion} as $(l+2)(l-1)/2$ which scales polynomially with $l$ inputs.
Thus, OQSS with $l$ inputs is expected to prepare an arbitrary state up to $\ket{n_{\rm max}}$ with
\begin{equation}
n_{\rm max}={(l+2)(l-1)}/{2}-1. \label{eqnmax}
\end{equation}

%\\\\\\\\\\\\\\\\\\\\\\\\\\\\\\\\\\\\\\\\\\\\\\\\\\\\\\\\\\\\\\\\\\\\\\\\\\\\\\\\\\\
%\\\\\\\\\\\\\\\\\\\\\\\\\\\\\\\\\\\\\\\\\\\\\\\\\\\\\\\\\\\\\\\\\\\\\\\\\\\\\\\\\\\
{\it Time complexity of the conventional method.}---
The problem of time complexity mainly comes from the calculation of a loop hafnian, which is a matrix function that counts the number of perfect matchings of weighted graphs with loops~\cite{quesada2019franck,quesada2019simulating}.
We here consider that an $l$-mode Gaussian state $\rho$, and the $i$-th mode for input and output contains $n_i$ and $m_i$ photons.
As introduced in Refs.~\cite{quesada2019franck,quesada2019simulating}, a loop hafnian appears in the Fock matrix elements of a Gaussian state to obtain the output of the circuit, and the Fock matrix elements are given by 
$\braket{{\bm m}|\rho|{\bm n}} \propto {{\rm lhaf}}(\tilde{\bm A}),$
where lhaf is a loop hafnian, and $\tilde{\bm A}$ is a square matrix of dimension $D=\sum_{i=1}^{l}(n_{i}+m_{i})$ with ${\bm n}=(n_1,\dots,n_l)$ and ${\bm m}=(m_1,\dots,m_l)$.
As shown in Ref.~\cite{quesada2019simulating}, the number of steps to calculate a loop hafnian is obtained by the smaller one of two values of
\begin{equation}
O(lA_{p}G^{l}_{p}), \hspace{10pt} O(l^2d^2d^{l}),  \label{eqloop1}
\end{equation}
where $A_{p}$, $G_{p}$, $d$ are the arithmetic means, geometric means, and a chosen truncated dimension for the output Hilbert space, respectively. $A_{p}$ and $G_{p}$ are given by $A_{p}=({1}/{l}) \sum_{i=1}^{l}(n_{i}+1)$ and $G_{p}=\left\{ \prod_{i=1}^{l} (n_{i}+1)\right\}^{1/l}$, respectively~\cite{quesada2019franck,quesada2019simulating}.
Equation~(\ref{eqloop1}) means that the computational time scaling for the conventional non-Gaussian state preparation scales exponentially with $l$ inputs.

%\\\\\\\\\\\\\\\\\\\\\\\\\\\\\\\\\\\\\\\\\\\\\\\\\\\\\\\\\\\\\\\\\\\\\\\\\\\\\\\\\\\
\begin{figure}[t]
\centering \includegraphics[angle=0, scale=1]{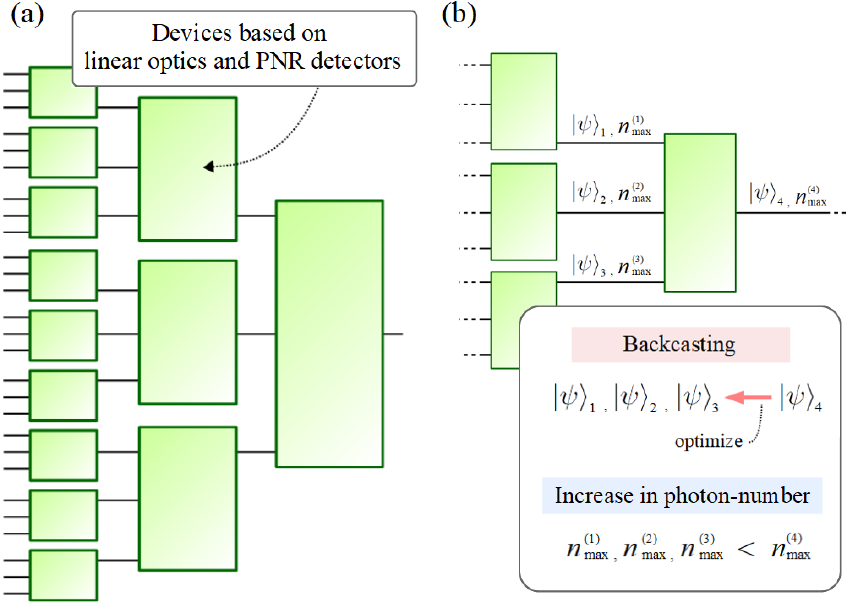} 
\caption{Concept of our scheme. (a) Proposed OQSS using a backcasting approach, where the circuit is divided into sublayers.
(b) Divided circuits are simulated backwards from final to first layers. 
 }
\label{fig2}
\end{figure}
%\\\\\\\\\\\\\\\\\\\\\\\\\\\\\\\\\\\\\\\\\\\\\\\\\\\\\\\\\\\\\\\\\\\\\\\\\\\\\\\\\\\

{\it Concept of our scheme.}---
The backcasting approach is the key to circumvent the time complexity. 
The conventional OQSS has a single circuit, while the proposed OQSS consists of multiple and layered circuits, as shown in Fig.~\ref{fig2}(a).
In the layered circuits, the truncated photon number of states progressively increases in each layer so that the truncated output photon number in the final layer becomes $n_{\rm max}$.
To determine the circuit parameters for each circuits, a backcasting approach is employed: it begins with a parameter estimation so that the circuit in the final layer generates the target state. Then, we optimize parameters $backwards$ from final to first layers, and determine the circuit parameters in the first layer.

In Fig.~\ref{fig2}(b), for instance, one component of the layered circuit prepares the state $\ket{\psi}_{4}$ using 3 inputs $\ket{\psi}_{i}$ ($i$=1,2,3), where $\ket{\psi}_{4}$ and $\ket{\psi}_{i}$ are described as Eq.~(\ref{eqfock}) with ${n_{\rm max}^{(4)}}$ and ${n_{\rm max}^{(i)}}$, respectively.
After conditioning on an appropriate pattern of the detected photon numbers and circuit parameters, ${n_{\rm max}^{(4)}}$ becomes larger than ${n_{\rm max}^{(i)}}$, where parameters are determined by considering $\ket{\psi}_{4}$ as the target for the circuit.
By repeating this procedure, we can obtain the target with a large $n_{\rm max}$ by using the set of small and simple circuits.

%=========================================
\begin{figure*}[t]
\centering \includegraphics[angle=0, scale=2]{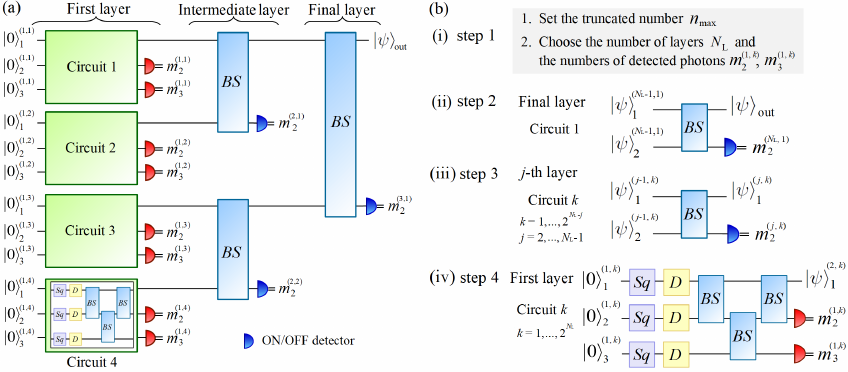} 
\caption{Proposed OQSS. (a) The proposed OQSS for a specific case with a single intermediate layer and 3 inputs for each of the circuits in the first layer. (b) The description of the whole procedure to determine the circuit parameters. (i) Step 1. (ii) Step 2. (iii) Step 3. (iv) Step 4.} 
\label{fig3}
\end{figure*}
%=========================================

{\it Specific case of our scheme.}---
Specifically, let us consider the case of 3 layers. Figure~\ref{fig3}(a) shows the schematic view of the proposed OQSS.
In the first layer, 3 input states in each of the circuits are combined by linear optics, and measured by PNR detectors except for the output from the $k$-th circuit in the first layer, $\ket{\psi}_{1}^{(2,k)}$.
Then, the output $\ket{\psi}_{1}^{(2,k)}$ becomes the input for the next layer.
The procedure for the first layer succeeds, when all detectors in the first layer count the predefined photon numbers, $m_{i}^{(1,k)}$.
To limit the complexity, % of the numerical simulation, 
the detected photon number in every detector and the number of inputs in the first step are set to at most 4. %, i.e. $m_{2}^{(1,k)}$, $m_{3}^{(1,k)}\leq 4$.

Then, in the intermediate layer, two inputs are coupled by a single beam splitter, and one of the states is measured by the on/off detector which can identify whether there is (a) photon(s) or not.
The generation of the states succeeds when all detectors in the first intermediate layer count the predefined photon numbers $m_{i}^{(2,k)}=0$~\cite{noteDetector}.
When the number of the intermediate layers $\geq$ 2, intermediate states are repeatedly generated in subsequent layers in a similar manner.
In the final layer, we obtain the output state, $\ket{\psi}_{1}^{(N_{\rm L},1)}$, approximated by Eq.~(\ref{eqfock}).

When considering $m_{2}^{(j,k)}$=0 $(2\leq j\leq N_{\rm L})$, the truncated photon number of the output, $n_{\rm max}$, is given by the sum of those of the outputs in the first layer as
\begin{equation}  
n_{\rm max} = \sum_{k=1}^{2^{N_{\rm L}-1}} n_{\rm max}^{(1,k)}=\sum_{k=1}^{{2^{N_{\rm L}-1}}}\left\{
 \sum_{i=2}^{3}{m_{i}^{(1,k)}} \right\}
, \label{nmax}
\end{equation}
where $N_{\rm L}$, ${m_{i}^{(1,k)}}$, and $n_{\rm max}^{(1,{k})}$ are the number of layers, the detected number of photon in the $i$-th input mode in the $k$-th circuit of the first layer, and the truncated photon number of the output in the $k$-th circuit of the first layer, respectively.

Although the numbers of input modes and $m_{2}^{(j,k)}$ for the $j$-th layer  $(2\leq j\leq N_{\rm L})$ are not necessarily two and zero, respectively, there are three reasons for doing this. First, the success probability for 2 inputs is larger that for more than 3 inputs.s
%, which is effective in improving the overall success probability when using a quantum memory, as described below.
Second, setting $m_{2}^{(j,k)}$ = 0 maximizes the truncated photon number $n_{\rm max}$ for the output. Third, using on/off detectors for measurements $m_{2}^{(j,k)}$ = 0 is experimentally more feasible than the use of PNR detectors.

Figure.~\ref{fig3}(b) shows the summary of the proposed algorithm consisting of 4 steps.
In step 1 (Fig.~\ref{fig3}(b)(i)), we set the truncated photon number for the target, $n_{\rm max}$.  
Then we set the number of the layer $N_{\rm L}$ and the detected photon numbers $m_{2}^{(j,k)}$ so that the values satisfy Eq.~(\ref{nmax}). From the next step, the circuit parameters are obtained by the optimization. In step 2 (Fig.~\ref{fig3}(b)(ii)), coefficients of two inputs and parameters for a beam splitter coupling are optimized so that the target GKP qubit in the mode 1 is generated from two inputs, $\ket{\psi}_{1}^{(N_{\rm L}-1,1)}$ and $\ket{\psi}_{2}^{(N_{\rm L}-1,1)}$, with no photon detection (i.e. $m_2^{(N_{\rm L},1)}=0$).
In step 3 (Fig.~\ref{fig3}(b)(iii)), we obtain the parameters repeatedly up to the second layer after replacement.
In step 4 (Fig.~\ref{fig3}(b)(iv)), finally, we determine the circuit parameters in the first layer.

Let us consider the generality of the possible, targeted outputs.
In the $k$-th circuit of the first layer, the circuit can generate an arbitrary state up to $n_{\rm max}^{(1,k)}= (l+2)(l-1)/2-1$ from Eq.~(\ref{eqnmax}). 
Then $n_{\rm nmax}$ would be conjectured as in Eq.~(\ref{nmax}).
Thus, the proposed OQSS has the potential to generate an arbitrary state up to
\begin{equation}
n_{\rm max}=\sum_{k=1}^{N_{\rm first}}\left\{\frac{(l_{k}+2)(l_{k}-1)}{2}-1\right\}, \label{eqinde2}
\end{equation}
where $N_{\rm first}$ is the number of circuits in the first layer, and $l_{k}$ is the number of inputs for the $k$-th circuit in the first layer.
For the specific circuit described in Fig.~\ref{fig3}, $N_{\rm first}$ corresponds to $2^{N_{\rm L}-1}$ in Eq.~(\ref{nmax}).

{\it Time complexity of the proposed scheme.}---In our method, the first layer mainly takes the number of steps to calculate a loop hafnian, where there are $2^{N_{\rm L}-1}$ circuits and the number of steps for each of the circuits is obtained by Eq.~(\ref{eqloop1}s).
The key of our scheme lies in choosing the low number of inputs.
When considering a positive integer, e.g. $l$=3 in Eq.~(\ref{eqloop1}), the number of steps is given by the smaller one of the two values of $O(3A_{p}G^{3}_{p}\ 2^{N_{\rm L}-1})$ and $O(9d^{5}\ 2^{N_{\rm L}-1})$.
More specifically, when we consider the case $n_{\rm max}^{(1,k)}=n_{\rm max}^{(1)}$ in Eq.~(\ref{nmax}), the numbers of steps become
$O(3A_{p}G^{3}_{p}\ n_{\rm max}/n_{\rm max}^{(1)})$ or 
$O(9d^{5}\ n_{\rm max}/n_{\rm max}^{(1)})$ 
using $n_{\rm max} = n_{\rm max}^{(1)}\ 2^{N_{\rm L}-1}$, where $d$ and $n_{\rm max}$ can be fixed to less than 100 and 50, respectively.

{\it GKP qubit.}---
We focus on the preparation of the GKP qubit~\cite{gottesman2001encoding}, which has two advantages towards optical FTQC with continuous variables~\cite{menicucci2014fault,fukui2018high,fukui2018tracking,baragiola2019all,walshe2019robust,pantaleoni2020modular,
walshe2020continuous,yamasaki2020cost,
pantaleoni2021subsystem,fukui2021efficient,bourassa2021blueprint,larsen2021fault,tzitrin2021fault}: (1) Error tolerance. 
The GKP qubit can achieve the hashing bound of the additive Gaussian noise~\cite{fukui2017analog,fukui2018high} and protects against a photon loss~\cite{albert2018performance}. 
(2) Scalability. Only by a beam splitter coupling, the GKP qubits can be entangled within a cluster state, where a large-scale cluster has been realized experimentally in optics~\cite{asavanant2019generation,larsen2019deterministic}.

%\\\\\\\\\\\\\\\\\\\\\\\\\\\\\\\\\\\\\\\\\\\\\\\\\\\\\\\\\\\\\\\\\\\\\\\\\\\\\\\\\\\
\begin{figure}[t]
 \centering \includegraphics[angle=0, scale=0.85]{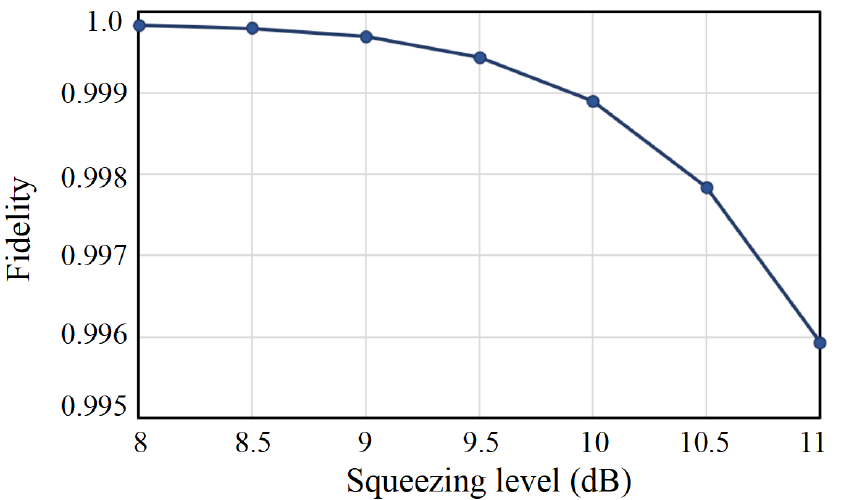} 
\caption{
The numerical results to prepare approximated GKP qubit with the truncated photon number $n_{\rm max}=32$. The fidelities $F_{\rm out}$ are plotted as a function of the squeezing level of the target.
}
\label{fig4}
\end{figure}
%\\\\\\\\\\\\\\\\\\\\\\\\\\\\\\\\\\\\\\\\\\\\\\\\\\\\\\\\\\\\\\\\\\\\\\\\\\\\\\\\\\\

In the numerical calculation, we target the 0 state for GKP codewords, $\ket{\overline{0}}$, where our proposed OQSS generates the approximated 0 state $\ket{\widetilde{0}}$ approximated by Eq.~(\ref{eqfock}), so as to be close to $\ket{\overline{0}}$.
Here we introduce the parameter $squeezing$ $level$ used for the threshold required for FTQC, and around 10 dB is often used for the threshold~\cite{fukui2018high,fukui2019high,noh2020fault,yamasaki2020polylog}.
To realize a high fidelity between $\ket{\overline{0}}$ and $\ket{\widetilde{0}}$, i.e. $|\braket{ \widetilde{0}|\overline{0}}|^2$, a large $n_{\rm max}$ is required as the squeezing level becomes larger.
We can roughly estimate the fidelity for $n_{\rm max}$ by considering $|\braket{\overline{0}_{n_{\rm max}}|\overline{0}}|^2$, where $\ket{\overline{0}_{n_{\rm max}}} \propto \sum_{i=0}^{n_{\rm max}}{g_{i}}\ket{i}$ and $g_{i}$ are coefficients of Fock basis states of $\ket{\overline{0}}$.
For example, $|\braket{\overline{0}_{n_{\rm max}}|\overline{0}}|^2$ with 10 dB, $\sim$99, $\sim$99.9, and $\sim$99.99~\%, corresponds to $n_{\rm max}=24$, 32, and 42, respectively.

{\it Numerical results.}---
%Our aim here is to show the realization of the efficient simulation of preparing the GKP qubit for FTQC.
%For this purpose, 
To prepare the GKP qubit for FTQC, we target $\ket{\overline{0}}$ around 10 dB, and $n_{\rm max}$ is set to 32.
For $n_{\rm max}=32$, %the setup to provide $n_{\rm max}=32$, 
we adopt a condition that sets the number of inputs for each circuit in the first layer to 3, the detected photon number for both detectors in the first layer is 2, and the number of layers is 4. %(i.e. the number of intermediate layers is 2).
This condition ensures the state preparation with a high fidelity, since we would access the number of independent coefficients up to the Fock state $\ket{32}$ %$\ket{n_{\rm max}}=\ket{32}$ 
using Eq. (\ref{eqinde2}) with $N_{\rm first}=8$ and $l_{k}=3$.

We evaluate the fidelity between the target and output states, $F_{\rm out}$, and the success probability of the output, $P_{\rm suc}$. To verify $F_{\rm out} \approx 99.9\%$, we numerically calculate the fidelity, $F_{\rm out}$=$\braket{ \overline{0}|\rho_{\rm out} | \overline{0}}$ with $\rho_{\rm out}=\ket{\psi}_{\rm out}\bra{\psi}_{\rm out}$, where parameters are optimized using Python modules the Walrus and Strawberry Fields~\cite{gupt2019walrus,killoran2019strawberry,tzitrin2020progress}.
In Fig.~\ref{fig4}, the fidelities for $n_{\rm max}=32$ are plotted as a function of the squeezing level of the target GKP qubit. The numerical results show that the fidelities are larger than 99.9\% with the squeezing smaller than $\sim $10 dB.
If a higher fidelity is required, corresponding to a squeezing value greater than 10 dB, we just need to increase $n_{\rm max}$.
The important feature of our method is that the simulation with larger $n_{\rm max}$ such as $n_{\rm max}=100$ can be also implemented in a realistic computational time, since the computational time for the first layer is not changed and the increase of the number of layers scales as polynomial time.
For the success probability, we obtained the success probability $P_{\rm suc}\approx10^{-29}$ for $n_{\rm max}=32$. To improve $P_{\rm suc}$, we could use the quantum memory (See supplemental materials for $F_{\rm out}$, $P_{\rm suc}$, and quantum memory).

{\it Discussion and conclusion.}---
We have developed an efficient way to simulate the non-Gaussian state preparation via the OQSS.
Our innovation can considerably reduce the simulation time on a classical computer to prepare an arbitrary non-Gaussian state with a large truncated photon number by employing a backcasting approach.  
As a specific example, we numerically showed that our scheme can simulate the preparation of the GKP qubit with a fidelity as high as required for FTQC in polynomial time.
Furthermore, the proposed OQSS offers the elimination of an experimental requirement for PNR detectors.
Conventionally, it has been assumed that a non-Gaussian target with a high fidelity and a large truncated number might considerably increase the number of detected photons and hence require unfeasible PNR detectors.

So most importantly from a practical point of view, our results show that the detected photon number by each detector is at most 2 which has been demonstrated~\cite{lita2008counting,fukuda2011titanium,endo2021quantum}.
Apart from FTQC, the OQSS allows us to prepare arbitrary non-Gaussian states which are an indispensable resource for many quantum technologies. 
Thus, our scheme can play a crucial role in realizing many applications for optical quantum information processing.

Finally, let us mention several directions for further investigations. First, some parameters, such as $N_{\rm L}$, $m_i^{(j,k)}$, and the number of modes in each circuit, are still predefined in our current approach. Thus, a generalized method to also find the circuit parameters that achieve a certain target fidelity by incorporating all such parameters would be a very useful extension of our work. Second, we could further evaluate the performance of our method when imperfections of the OQSS, such as photon loss, inefficient detectors, and anti-squeezing effects~\cite{walshe2019robust}, are taken into account~\cite{noteImperfection}. Lastly, this work focuses on the preparation of single-mode states. Nonetheless, the OQSS can potentially also prepare multi-mode states. It is an interesting open question what kind of multi-mode states the OQSS can generate~\cite{noteMultimode}.

{\it Acknowledgements.}---
We thank Kenji Yamanishi, Atsushi Nitanda, and Taichi Kiwaki for useful discussions.
This work was partly supported by JST [Moonshot R$\&$D][Grant No. JPMJMS2064][Grant No. JPMJMS2061], Japan Society for the Promotion of Science (JSPS) KAKENHI (grant 18H05207) (grant 20K15187), UTokyo Foundation, and donations from Nichia Corporation. 
P.v.L. acknowledges financial support from BMBF via QLinkX and from BMBF/EU-Quantera via ShoQC.
M.E. acknowledges supports from Research Foundation for Opto-Science and Technology.

\bibliography{ref.bib}

\begin{thebibliography}{10}
\bibitem{noteDetector}
``The performance of the {ON/OFF} detector, which discriminates the detected
  photon numbers between zero and nonzero, might require high-quality photon
  number discrimination capability comparable to that of a {PNR} detector.''
\end{thebibliography}
\bibliographystyle{IEEEtran}

\end{document}